# Modeling the direct effect of solvent viscosity on the physical mass transfer in a packed column

Zhijie Xu[1], Rajesh Kumar Singh[1] and Jie Bao[2]

[1]Physical and Computational Sciences Directorate,
Pacific Northwest National Laboratory, Richland, WA 99352, USA

[2]Energy and Environment Directorate,
Pacific Northwest National Laboratory, Richland, WA 99352, USA

**ABSTRACT**

Post-combustion carbon capture via solvent absorption in a packed column can be a promising technology for mitigating greenhouse gas emissions. The mass transfer between gas and liquid solvent phases plays a critical role in determining the packing height of the packed column. Recent experiments [1] involving physical mass transfer in a packed column have distinguished the direct and indirect impacts of viscosity ($\mu_L$) on the mass transfer coefficient ( $k_L$). The direct impact of viscosity on the mass transfer coefficient has been found to be significantly higher for a packed column ($k_L \sim \mu_L^{-0.38}$) than a wetted wall column (WWC) ($k_L \sim \mu_L^{-0.17}$). To explain the deviation from the well-controlled uniform film flow in a WWC, we propose a mechanism involving the unavoidable wavy film flow and eddy-enhanced mass transfer in a packed column. A mathematical analysis based on fundamental flow and mass transfer equations is presented to quantify this deviation. The correlation for mass transfer coefficient derived from the mathematical analysis is found to match the corresponding one derived from the experiments for a packed column. Multiphase flow simulations using the Volume of Fluid method also are conducted to verify the direct impact of viscosity on the mass transfer coefficient for both uniform and wavy film flows. The mass transfer coefficient calculated from simulations shows steeper variation with viscosity for wavy film compared to the uniform film, further confirming our proposed theory. A similar relation ($k_L \sim \mu^{-0.38}$) for wavy film flow in a packed column is observed from both theoretical analyses, experimental observations, and computer simulations.

## 1. Introduction

Solvent absorption by countercurrent flow in a packed column is a promising technology for mitigating carbon dioxide ($CO_2$) emissions from power plants [2-4]. The structured/random packing used in the packed column provides a large surface area for mass transfer while minimizing pressure drop across the column [5]. Before deploying this technology to industry, its performance and scalability must be investigated [6]. Accordingly, hydraulic characteristics and mass transfer between flue gas and liquid solvent phases in the packing system need to be better understood for optimal packing design. It is a challenging problem because of the large number of factors that influence the hydrodynamics and mass transfer behavior, such as solvent properties, surface characteristics, packing design, and liquid loading [7]. In this view, understanding the local hydrodynamics and mass transfer behavior is important as it reflects the liquid flow pattern within the column that significantly influences the overall mass transfer rate between phases. In addition, the mass transfer between phases governs the packing height of the column.

A number of experimental and theoretical studies have been conducted to explain the mechanism for mass transfer in structured packings. Various theories, including two-film theory [8], penetration theory [9], surface renewal theory [10], and boundary layer theory, are applied to explain the mass transfer in a packed column. Most of the models for physical mass transfer are based on two-film theory, which only considers the molecular diffusion in stagnant films that form each side of the gas-liquid interface [8]. On the other hand, Higbie's penetration theory [9] considers the residence time of a fluid element at the interface. It suggests an elementary idea to account for turbulence in mass transfer phenomena. Subsequently, correlation for the mass transfer coefficient is derived in terms of various dimensionless numbers, such as Reynolds, Schmidt, Froude, or Galileo number. Previously, the wetted wall column (WWC) concept was adopted to predict mass transfer based on the assumption of well-defined geometry and precise surface area in the structured packings. However, recent investigations have shown complex flow behavior inside the packed column along with the presence of different flow morphologies. As a result, the mass transfer prediction significantly deviates from the WWC. The mass transfer between phases is dictated by many factors. Surface tension does not affect mass transfer in the turbulent wavy film flow [11]. On the other hand, viscosity is one of the critical factors affecting the mass transfer phenomenon. A large value of solvent viscosity can lead to a significant reduction in the mass transfer coefficient. High solvent viscosity can slow diffusion of gas through the gas-liquid interface (indirect effect) and modify the local hydrodynamics by generating a more stabilized liquid film on the packing surface (direct effect). As a result, mass transfer across the interface can be significantly reduced.

The effect of viscosity on the mass transfer coefficient has been investigated in detail over the last few decades. Rochelle and his coworkers [1, 12-15] have performed extensive experimental studies examining the performance evaluation of structure/random packings. The effects of various factors impacting the mass transfer area [13-15] and mass transfer coefficients between phases [1, 12] also have been extensively investigated. Song et al. [1, 12] have experimentally investigated the physical mass transfer for various carbon capture packings. The effect of viscosity on the overall mass transfer coefficient has been explained in two parts: 1) direct contribution due to hydrodynamics and 2) indirect impact that accounts for solvent diffusivity. An experimental campaign designed to study these factors could be expensive and time consuming. Conversely, computational fluid dynamic (CFD) simulations may be leveraged in the design process to achieve economic and efficient solutions while providing important insights into the physics. Notably, CFD does not replace the need for measurements, but it may diminish the amount of experimentation required and complement the setup and analysis. In addition, CFD steadily has been gaining acceptance for studying the flow characteristics in structured packings over time [16-20].

Haroun et al. [21, 22] and Marschall et al. [23] have developed the single fluid formulation for mass transfer between two phases. It is capable of simulating species transfer across interfaces having arbitrary morphology using interface capturing methods, such as the Volume of Fluid (VOF) method. To capture the concentration jump across the interface, both approaches utilized Henry's law with a constant coefficient for thermodynamic equilibrium of chemical species. This approach also is employed [24-27] for various chemical processes where mass transfer is important. Nieves-Remacha et al. [25] use this approach to simulate the hydrodynamics and mass transfer phenomenon in advanced-flow reactor technology, an alternative to scale up continuous flow chemistries from micron to millimeter scales. Hu et al. [28] have conducted a VOF simulation similar to the method suggested by Haroun et al. [22] for mass transfer in a falling wavy film. They characterize the $CO_2$ absorption mass transfer in vorticity in the vicinity of the interface and the local $CO_2$ concentration. In the same context, Singh et al. [7, 29] have conducted VOF simulations for film and rivulet flow over an inclined plate for the shape gas-liquid interface and flow morphology. Extensive multiphase flow simulations also have been conducted that account for the factors affecting rivulet shape and stability. Note that the shape of the interface significantly affects the mass transfer across it. Sebastia-Saez et al. [30] have investigated physical mass transfer in a countercurrent flow over an inclined plate. Mass transfer due to physical absorption shows non-monotonic variation with increased liquid load. The maximum value of mass transfer is observed at the fully wetted film. Recently, Chao et al. [26, 27] have applied a similar approach for physical and reactive mass transfer for $CO_2$ capture in a WWC, where they show the advantage of CFD modeling over the standard two-film theory. In addition, CFD modeling can account for the local flow hydrodynamics, such as wavy interface, and variation in film thickness in $CO_2$ capture. Haroun et al. [22] shows that mass transfer between phases strongly depends on the packing design. A sheet of triangular channel shows higher mass transfer than a flat surface. However, their problem setup was limited to concurrent gas-liquid flow.

In a recent experimental study by Song et al. [1], the direct effect of viscosity on the mass transfer coefficient is shown to be significantly higher than the corresponding one for a WWC. The mass transfer coefficient for a WWC can be expressed as:

$$k_L = 1.15 \cdot u_L^{0.333} \nu_L^{-1/6} g^{1/6} \left( \frac{a^{1/3}}{L^{1/2}} \right) D_L^{1/2}, \tag{1}$$

where $u_L$ is solvent velocity at the inlet, $\nu_L$ is the kinematic viscosity, $g$ is the gravitational acceleration, $D_L$ is the diffusivity of the transport species in the solvent phase, $a$ is the solvent inlet size, and $L$ is the WWC height. This correlation has deep origins from solving the governing equations for laminar flow of a flat film over a vertical plate with coupled convective and diffusive mass transfer and the penetration theory [31, 32]. On the other hand, a similar correlation also can be obtained based on extensive experiments on the packed column [1]:

$$k_L = 0.12 \cdot u_L^{0.565} v_L^{-2/5} g^{1/6} a_p^{-0.065} D_L^{1/2}, \qquad (2)$$

where $a_p$ is specific area of the packings presented as $m^2/m^3$. Equation (2) clearly shows a higher impact of solvent viscosity on the mass transfer coefficient for a packed column with an exponent of -2/5 as compared to -1/6 for the WWC (Eq. (1)). In addition, a larger dependence of mass transfer on the solvent velocity ($u_L$) also can be identified for the packed column with an exponent of 0.565 compared to 0.33 in Eq. (1). Of note, the flow in a packed column is more complicated compared to WWCs because of various factors, including the intermittent nature of the liquid distributor, surface characteristics of the packing material, and turbulent nature of flows due to design and arrangement of packing units. As a result, flow in the packed column significantly deviates from a well-controlled laminar flow found in the WWC. Subsequently, it may contribute to the enhanced dependency of mass transfer on the viscosity [1]. However, unlike correlation for WWC (Eq. (1)) that is corroborated by a rigorous and quantitative theory, the exact mechanism is not fully understood for the packed column, and associated quantitative analysis is still lacking. In this context, an extensive theoretical study with numerical simulations will provide significant insights and quantitative explanations of the physics behind the correlation in Eq. (2), as well as additional insights into the design optimization of a packed column.

In this paper, an analytical derivation for the dependence of mass transfer on viscosity is presented in Section 2. Section 3 briefly describes the formulation for VOF simulations followed by a description of the multiphase flow simulation setup and meshing of computational domain. The simulation results and comparison for the mass transfer coefficient are presented in Section 4 followed by a conclusion.

## 2. Mathematical Formulation of the Physical Mass Transfer

We consider a wavy film with rolling waves falling over a vertical flat plate due to gravity with the physical mass transfer of gas species across the gas-liquid interface. Figure 1 shows the schematic of the problem, where rolling waves can be characterized by an amplitude *A*, wavelength λ, and wavespeed $\boldsymbol{u_w}$. As a result, the thickness profile $\delta(\boldsymbol{x},\boldsymbol{t})$ of the wavy film varies vertically in the falling direction. The *x*-axis is directed downward in the streamwise direction (see Figure 1), and the *y*-axis is perpendicular to the plate. The governing (continuity and momentum conservation) equations dictating the falling film are given as:

$$\frac{\partial u}{\partial x}+\frac{\partial v}{\partial y}=0\,, \tag{3}$$

$$\frac{\partial u}{\partial t}+u\frac{\partial u}{\partial x}+v\frac{\partial u}{\partial y}=-\frac{1}{\rho_L}\frac{\partial P}{\partial x}+\frac{\mu_L}{\rho_L}\left(\frac{\partial^2 u}{\partial x^2}+\frac{\partial^2 u}{\partial y^2}\right)+g\,, \tag{4}$$

$$\frac{\partial v}{\partial t}+u\frac{\partial v}{\partial x}+v\frac{\partial v}{\partial y}=-\frac{1}{\rho_L}\frac{\partial P}{\partial y}+\frac{\mu_L}{\rho_L}\left(\frac{\partial^2 v}{\partial x^2}+\frac{\partial^2 v}{\partial y^2}\right), \tag{5}$$

where *u* and *v* are the vertical and horizontal components of the velocity, $\rho_L$ is the liquid density, $\mu_L$ is the dynamic viscosity, and *P* is the pressure. A parabolic velocity profile in the streamwise direction for *u* may be established across the film, which is based on the exact solution of a steady and fully developed laminar film flow [33]:

$$u\left(x,y\right)=\frac{3}{2}\frac{q}{w}\left(\frac{2y}{\delta^2\left(x\right)}-\frac{y^2}{\delta^3\left(x\right)}\right). \tag{6}$$

Here, *w* is the width of film along the direction perpendicular to the x-y plane, and *q* is the flow rate,

$$q=w\int_0^{\delta(x)}u\left(x,y\right)dy=w\cdot u_{av}(x)\delta\left(x\right), \tag{7}$$

where $u_{av}(x)$ is the average vertical velocity at a location *x*. Using Eq. (6), the gradient of vertical velocity *u* in the *x* direction can be related to the local film thickness δ(x),

$$\frac{\partial u}{\partial x}=\frac{\partial u}{\partial \delta}\frac{\partial \delta}{\partial x}=-\frac{3}{2}\frac{q}{w}\frac{y}{\delta^4}\left(4\delta-3y\right)\frac{\partial \delta}{\partial x}. \tag{8}$$

The gradient of lateral velocity *v* along the *y* direction at the interface can be computed using the continuity equation (3)

$$\left.\frac{\partial v}{\partial y}\right|_{y=\delta(x)}=-\left.\frac{\partial u}{\partial x}\right|_{y=\delta(x)}=\frac{3}{2}\frac{q}{w}\frac{1}{\delta_0^2}\frac{\partial \delta}{\partial x}, \tag{9}$$

where $\delta_0$ is the average film thickness that can be written as

$$\delta_0=\left(\frac{3\mu_L^2\,\mathrm{Re}}{\rho_L^2 g}\right)^{1/3} \tag{10}$$

according to the Nusselt theory [34]. Here, Re is the film Reynolds number defined as,

$$\mathrm{Re}=\frac{\rho_L q}{\mu_L w}=\frac{q}{\nu_L w}=\frac{u_L a}{\nu_L}, \tag{11}$$

where *a* is the size of the solvent inlet, $u_L$ is the solvent velocity at inlet, and $\nu_L$ is the kinematic viscosity of the solvent. For a turbulent wavy film flow, transport near a gas-liquid interface is governed by eddies whose length and velocity scales are characterized by bulk turbulence [11, 35]. According to the mixing length model for eddy diffusion [36], the eddy diffusivity stemming from the local turbulent flow in the direction normal to the vertical plate can be written as:

$$D_{Le}=C_1\cdot\left|\left.\frac{\partial v}{\partial y}\right|_{y=\delta(x)}\right|\left(\Delta y\right)^2. \tag{12}$$

Here, $C_1$ is a proportional constant, Δy is the penetration depth of gas species into the solvent during a single wave period $T=\lambda/u_w$ (shown in Figure 1). The penetration depth Δy depends on the wave frequency *f* and can be related to the diffusivity of that species in liquid phase ($D_L$) and the frequency ($f$) of rolling waves,

$$\Delta y=\sqrt{D_L/f}=\sqrt{D_L\,\lambda/u_w}\,. \tag{13}$$

Substitution of Eqs. (9) and (13) into Eq. (12) leads to the expression for eddy diffusivity $D_{Le}$:

$$D_{Le} = \left[ 0.72 C_1 \frac{\partial \delta}{\partial x} f \left( \frac{\rho_L g^2}{\mu_L} \right)^{1/3} \mathrm{Re}^{1/3} \right] D_L . \tag{14}$$

To determine $D_{Le}$, the thickness gradient $\partial\delta/\partial x$ and the wave frequency *f* need to be modeled explicitly. We assume that an arbitrary fluid element at the gas-liquid interface follows a random walk motion along the *y* direction (shown in Figure 1) due to the momentum transfer. The velocity of that fluid element over a single period *T* of the rolling wave can be written as:

$$V_y = \sqrt{\nu_L \frac{u_w}{\lambda}} = \sqrt{\nu_L f} , \tag{15}$$

where, again, $\nu_L$ is the kinematic viscosity; λ is the wavelength; $u_w$ is the wave speed; and $f = u_w/\lambda$ is the wave frequency. Concurrently, the same fluid element also will experience downward motion along the *x* direction because of the gravity and vertical velocity gained during the same period *T* can be written as:

$$V_x = g \frac{\lambda}{u_w} = \frac{g}{f} . \tag{16}$$

The gradient of film thickness along the *x* direction ($\partial\delta/\partial x$) can be approximated using the wave amplitude *A* and wavelength λ as:

$$\frac{\partial \delta}{\partial x} \approx \frac{2A}{\lambda} . \tag{17}$$

*A* and λ can be related to the distance that the fluid element travels in vertical and lateral directions in a single rolling wave period *T*. These two quantities should be proportional to the speed in both directions and can be written as:

$$\frac{1}{2} \frac{\partial \delta}{\partial x} \approx \frac{A}{\lambda} = C_2 \frac{V_y}{V_x} , \tag{18}$$

where $C_2$ is a dimensionless constant on the order of unity. To verify this relation, we use the results obtained from the nonlinear solution of the wavy film falling over a vertical plate by Shkadov [37]. The various optimum parameters of the wave regime, such as wavelength, wave propagation speed, and frequency, can be computed as a function of the Reynolds and Kapitza numbers (*Ka*), where *Ka* incorporates the effect of surface tension:

$$Ka = \frac{\sigma}{\rho_L}\left(\frac{\rho_L^4}{g\mu_L^4}\right)^{1/3} = \frac{\sigma}{\rho_L}\left(\frac{1}{g\nu_L^4}\right)^{1/3}, \tag{19}$$

where $\sigma$ is the surface tension. In Figure 2, we plot the variation of the computed value of $A/\lambda$ with $V_y/V_x$ at three given Reynolds numbers with the Kapitza number varying between 400 and 4000. It can be observed that the variation of $A/\lambda$ with $V_y/V_x$ remains unchanged for all Reynolds and Kapitza numbers studied. Furthermore, it clearly shows the linear scaling relation as $A/\lambda = C_2 V_y/V_x$ with $C_2 \approx 0.67$. The numerical solutions ([37]) are in good agreement with our projection (Eq. (18)) and confirm that the ratio between wave amplitude and wavelength is proportional to the ratio of two velocity components gained during a single wave period *T*.

Next, the ratio $A/\lambda$ can be plotted as a function of Reynolds and Kapitza numbers using the same solution methods [37],

$$\frac{A}{\lambda} = C_3 Ka^{-n}\,\mathrm{Re}^m. \tag{20}$$

Here, $C_3$ is another dimensionless constant. In Figure 3(a), we present the variation of $A/\lambda$ with Reynolds number *Re*. As expected, $A/\lambda$ increases with an increased value of *Re*. Conversely, $A/\lambda$ decreases with an increased value of *Ka,* which is shown in Figure 3(b) for two different Reynolds numbers. Physically, it can be understood by the fact that the increased value of the Kapitza number leads to a reduction in film thickness and an increase in wavelength. Subsequently, fluctuations in film thickness also decrease, causing a lower value of $A/\lambda$. Both observations also are confirmed in the numerical simulations and will be explained in a later section (Figures 6 and 7). Accordingly, $A/\lambda$ decreases with the increased value of *Ka*. In addition, $A/\lambda$ shows nonlinear variation with *Re* and *Ka*. To find the scaling relation for the variation of $A/\lambda$ with *Re* and *Ka*, nonlinear regression is conducted, and a scaling relation $A/\lambda = C_3\,\mathrm{Re}^{0.78}\,Ka^{-0.47}$ is observed (shown in Figure 4) with $C_3 \approx 1/60$, with $m \approx 3/4$ *and* $n \approx 3/7$.

Finally, we can write these relations together as:

$$\frac{1}{2}\frac{\partial \delta}{\partial x} \approx \frac{A}{\lambda} = C_2 \frac{V_y}{V_x} = C_2 \frac{\sqrt{\nu_L}}{g} f^{3/2} = C_3 Ka^{-n}\,\mathrm{Re}^{m} \,. \qquad (21)$$

Substituting Eq. (21) into Eq. (14), the final expression for the eddy diffusivity for wavy film flow with rolling waves can be presented as:

$$\begin{aligned} D_{Le} &= 1.44\left[ C_1 \cdot C_2 \left( \frac{V_y}{V_x} \right)^{1/3} \mathrm{Re}^{1/3} \right] D_L \\ &= 1.44 C_1 \cdot C_2^{2/3} C_3^{1/3}\, \mathrm{Re}^{(m+1)/3}\, Ka^{-n/3} D_L \\ &= C\, \mathrm{Re}^{(m+1)/3}\, Ka^{-n/3} D_L \end{aligned} \qquad (22)$$

Here, $C = 1.44 C_1 C_2^{2/3} C_3^{1/3}$ is another empirical constant that aggregates three constants together. To derive the expression for effect of viscosity on the mass transfer coefficient in a wavy film flow with rolling waves, the molecular diffusivity $D_L$ in the WWC correlation (Eq. (1)) can be replaced by the eddy diffusivity $D_{Le}$. The difference in mass transfer between wavy film and uniform film flow primarily is assumed to be primarily due to the difference in diffusivity. It generates the following expression:

$$\begin{aligned} k_{Le} &= 1.15 u_L^{1/3} \left( \frac{\mu_L}{\rho_L} \right)^{1/6} g^{1/6} \left( \frac{a^{1/3}}{L^{1/2}} \right) D_{Le}^{1/2} \\ &= 1.15 u_L^{1/3} \left( \frac{1}{\nu_L} \right)^{1/6} g^{1/6} \left( \frac{a^{1/3}}{L^{1/2}} \right) C^{1/2} \left( \frac{u_L a}{\nu_L} \right)^{(m+1)/6} \left( \frac{\sigma}{\rho_L} \left( \frac{1}{\nu_L^4 g} \right)^{1/3} \right)^{n/6} D_L^{1/2} \\ &= 1.15 C^{1/2} u_L^{(3+m)/6} g^{(3-n)/18} \left( \frac{a^{(3+m)/6}}{L^{1/2}} \right) \left( \frac{\sigma}{\rho_L} \right)^{n/6} \nu_L^{-(6+3m+4n)/18} D_L^{1/2} \end{aligned} \,. \qquad (23)$$

The original correlation for WWC (Eq. (1)) can be fully recovered by substituting the value of the constant *C*=1, exponents *m* = -1, and *n* = 0 in Eq. (23). However, in the case of a packed column, these values can differ because of the turbulent wavy nature of flow. From the solutions of wavy film flow with rolling waves, we find the exact values of exponents as $m \approx 3/4$ *and* $n \approx 3/7$ that can be estimated from the scaling relations (Eq. (20)), shown in Figure 4.

Substituting the value of exponents of $m$ and $n$ into Eq. (22), the eddy diffusivity can be related to the Reynolds number (Re) and Kapitza number (Ka)

$$D_{Le} = C\,\mathrm{Re}^{7/12}\,Ka^{-1/7} D_L, \tag{24}$$

and the corresponding mass transfer coefficient ( $k_{Le}$ ) can be expressed as:

$$\begin{aligned} k_{Le} &= 1.15 C^{1/2} u_L^{5/8} g^{4/21} \left(\frac{a^{5/8}}{L^{1/2}}\right)\left(\frac{\sigma}{\rho}\right)^{-1/14} \nu_L^{-61/168} D_L^{1/2} \\ &= 1.15 C^{1/2} u_L^{0.625} g^{0.19} \left(\frac{a^{0.625}}{L^{0.5}}\right)\left(\frac{\sigma}{\rho}\right)^{-0.07} \nu_L^{-0.36} D_L^{1/2} \end{aligned} . \tag{25}$$

The preceding correlation is based on the coupling of hydrodynamic equations and eddy diffusion. It can be quantitatively compared to the correlation derived from extensive experiments for a packed column (Eq. (2), repeated here for convenience),

$$k_L = 0.12 \cdot u_L^{0.565} g^{0.17} a_p^{-0.065} \nu_L^{-0.4} D_L^{0.5}. \tag{26}$$

It is clear that the mass transfer coefficient for wavy film has significantly higher dependence on both the solvent flow rate and viscosity compared to the uniform film flow in a WWC (Eq. (1)). This dependence matches well with the correlation (Eq. (26)) developed from an experimental investigation for a packed column [1]. The expression presented in Eq. (25) confirms the functional form of the correlation derived from the experimental observations (Eq. (26)) with a slight difference in the exponent values. The proposed mechanism, i.e., enhanced diffusivity due to wavy film flow in a packed column, reasonably captures the deviation of physical mass transfer from the WWC.  Furthermore, the present expression for the mass transfer coefficient explores additional contributions from interfacial surface tension. Note that this term is missing in the experimentally derived correlation (Eq. (26)), which highlights the impact of surface tension on the hydrodynamics, thereby influencing the mass transfer phenomena. While it may be minor for an aqueous amine-based solvent, it can be more significant for a solvent with a higher value of interfacial surface tension.

Moreover, the mass transfer coefficient can be normalized by $D_L/(\nu_L^2/g)^{1/3}$, where the term $(\nu_L^2/g)^{1/3}$ is the viscous length scale. A similar method also has been applied earlier by Yih [35]. The normalized mass transfer coefficient $(k_{Le}^*)$ depends on several dimensionless numbers:

$$k_{Le}^* = \frac{k_L}{D_L}\left(\frac{\nu_L^2}{g}\right)^{1/3} = 1.15C^{1/2} \cdot \mathrm{Re}^{(3+m)/6}\,\mathrm{Sc}^{1/2}\mathrm{Ka}^{-n/6}\mathrm{Ga}^{-1/6}\,. \tag{27}$$

Here, $\mathrm{Sc}=\nu_L/D_L$ is the Schmidt number, and $\mathrm{Ga}=gL^3/\nu_L^2$ is the Galileo number. In the case of a WWC, the normalized mass transfer coefficient can be obtained by substituting the value as $C = 1, m = -1$, and $n = 0$ in Eq. (27):

$$k_{L(WWC)}^* = 1.15 \cdot \mathrm{Re}^{1/3}\,Sc^{1/2}\mathrm{Ga}^{-1/6}\,. \tag{28}$$

Therefore, enhancement in the mass transfer for a wavy film can be expressed as:

$$E = \frac{k_{Le}^*}{k_{L(WWC)}^*} = C^{1/2}\,\mathrm{Re}^{(1+m)/6}\,Ka^{-n/6}\,. \tag{29}$$

A greater enhancement in the mass transfer can be expected for a large Reynolds number and small Kapitza number because $m > 0$ and $n > 0$ for a wavy film flow with rolling waves.

## 3. Numerical Modeling

Multiphase flow simulations for mass transfer of a WWC and zigzag column (ZZC) are performed using the VOF method [38] to verify the analytical derivation of the physical mass transfer in the previous section. Physical mass transfer across the interface is modeled by a gas dissolution model based on the Rayleigh–Plesset equation available in Star-CCM$^+$. This method is fully explained in the Star-CCM$^+$ User Guide [39] with only a brief outline provided here. The governing equations for flow are as follows:

$$\nabla \cdot \mathbf{u} = 0\,, \tag{30}$$

$$\frac{\partial(\rho\mathbf{u})}{\partial t} + \nabla\cdot(\rho\mathbf{u}\mathbf{u}) = -\nabla p + \mu\nabla\cdot\left(\nabla\mathbf{u} + \left(\nabla\mathbf{u}\right)^T\right) + \rho\mathbf{g} + \mathbf{F}\,, \tag{31}$$

where $p$ is pressure and $g$ is the gravitational acceleration. The surface tension force, **F**, which produces a jump in the normal across the interface, is expressed as a singular body force. The terms $\rho$ and $\mu$ are the phase average density and viscosity, respectively. The surface tension force (**F**) appears at the gas-liquid interface and is implemented through the continuous surface force (CSF) model [40]:

$$\mathbf{F} = \sigma \frac{\rho \kappa \nabla \alpha}{\frac{1}{2}\left(\rho_g + \rho_l\right)}, \tag{32}$$

where $\sigma$ is the interfacial tension, $\kappa$ is the local curvature of the interface, $\alpha$ is the volume fraction of the liquid phase, and $\nabla\alpha$ can be related to the normal vector that is perpendicular to the interface. The curvature $\kappa$ is computed by the divergence of the unit normal $\hat{n} = \nabla \alpha / \|\nabla \alpha\|$ at the interface,

$$\kappa = \nabla \cdot \hat{n}. \tag{33}$$

The interface between the phases is captured by solving an additional transport equation (34), where the value of $\alpha$ is between 0 to 1,

$$\frac{\partial \alpha}{\partial t} + \mathbf{u} \cdot \nabla \alpha = 0. \tag{34}$$

Equation (34) is solved only for the liquid phase, and the volume fraction of the other phase is computed as $(1-\alpha)$.

Implicit transient flow simulations are conducted using Star-CCM$^+$ [39]. Velocity and pressure are coupled using the Pressure-Implicit with Splitting of Operators (PISO) algorithm [19]. The second-order upwind scheme is used in the spatial discretization of all equations. The geo-reconstruct scheme is used in the discretization of the interface. An implicit solution scheme is used in conjunction with an algebraic multigrid method (AMG) for accelerating convergence of the solver. Convergence of the solution is assumed when the sum of normalized residual for each conservation equation is less than or equal to $10^{-5}$.

We have investigated the influence of solvent viscosity on the mass transfer for both wavy film flow with rolling waves and uniform film flows in a WWC and a ZZC. The periodic wavy film is introduced via periodic perturbations at the solvent inlet. The WWC geometry is chosen from the previous study of Wang et al. [26, 27]. The ZZC is selected for its two-dimensional (2D) representation of structured packed columns. A similar approach also has been considered earlier for 2D representation of the structured packed column for investigating the hydrodynamics of countercurrent flow [18, 19]. Figure 5 depicts the schematic of the WWC and ZZC column designs and their corresponding parameters. Both columns are placed vertically and gravity acts in a downward direction. Dimension (height and width) of the WWC and ZZC geometries are kept the same to compare the mass transfer coefficient for both cases. The inlet and outlet of the gas and liquid phases have 1 mm widths. This inlet and outlet size is selected based on a previous WWC study [27].

The computational flow domain is developed and meshed in StarCCM$^+$. To accurately and efficiently capture the mass transfer phenomenon, the domain is discretized using very fine mesh. The mesh density proximate to the left wall where liquid falls appears finer than in the adjacent region for the surrounding gas. A more refined mesh also is required in the inflexion region near the zigzag channel apex to capture the flow behavior (see the dark region in the exploded view of the mesh domain in Figure 5(c)). Such simulations require a very small variable time step ($\Delta t \sim 10^{-6} - 10^{-5}$ sec) to satisfy the Courant – Friedrichs – Lewy (CFL) condition for stability with a Courant number = 0.50. As a result, flow simulations become quite expensive computationally.

Countercurrent flow simulations have been conducted for the multicomponent gas (air and nitrous oxide, $N_2O$) and amine-based solvents ($\rho_L = 1058\ kg/m^3$) as working fluids. The solvent viscosity is varied over a wide range ($\mu_L \sim 1 - 10\ cP$) to investigate its direct impact on the mass transfer between two phases. The gas phase has a density of ($\rho_g$)$1\ kg/m^3$ and viscosity ($\mu_g$) of $0.015\ cP$. The diffusivities of $N_2O$ in the air and solvent are specified as $D_G = 1.67 \times 10^{-5}\ m^2/s$ and $D_L = 1.3 \times 10^{-9}\ \frac{m^2}{s}$, respectively. The solvent enters the domain at the top (1) and exits the bottom (3) because of gravity. Both uniform (0.17 m/sec) and oscillating velocity profiles [$0.17(1 + 0.05 sin 2\pi\omega)\ m/sec$ and $\omega = 60Hz$] are prescribed at the solvent inlet to introduce the flat film flow and wavy film flow, respectively. On the other hand, uniform gas inlet velocity (0.049 m/sec) with $N_2O$ concertation of $13.31\ mol/m^3$ is specified at the bottom (see gas inlet (4) in Figure 5 (a and b). The outlet of both phases is specified as a pressure outlet with zero gauge pressure. The remaining boundaries, including zigzag sheet, are defined as a no-slip wall with a static contact angle ($\gamma$) of 70°. The benchmark results correspond to fixed liquid flow rates. In that case, a flow rate of $u_L = 0.17\ m/s$ and $u_G = 0.049\ m/s$ ensure a stable film flow with a smooth interface for a flat surface. The simulation cases are summarized in the following table:

| Viscosity | Wetted wall column (WWC) | | Zigzag column (ZZC) | |
|---|---|---|---|---|
| | Flat film flow with uniform flow rate | Wavy film flow with oscillating flow rate | Flat film flow with uniform flow rate | Wavy film flow with oscillating flow rate |
| 1 cP ~ 10 cP | UWWC | OWWC | UZZC | OZZC |

## 4. Results and Discussion

Analytical and numerical studies have been conducted for the mass transfer via physical absorption in a countercurrent gas-liquid flow. To explain its impact on the mass transfer coefficient, the flow simulations have been conducted for a range of solvent viscosity (1–10 cP) for uniform and oscillating inlet flow. In the simulations, we only consider the direct effect of solvent viscosity that affects the hydrodynamics. The indirect impact of viscosity on the mass transfer coefficient via viscosity-dependent diffusivity is not included, which involves the relations between diffusivity and viscosity of the solvents. The simulated results are presented in terms of dimensionless numbers: Reynolds, Kapitza, Schmidt, and Galileo numbers. These numbers already are defined in the previous section.

### 4.1 Film thickness

A thin film facilitates the mass transfer between the gas and liquid phases, and the mass transfer coefficient strongly depends on the film thickness. In addition, it also dedicates the heat transfer between the solvents and wall. In this context, computation of the film thickness is performed first. The effect of viscosity on the film thickness is presented in terms of Kapitza number, a dimensionless number that depends only on the solvent properties. Figure 6(a) shows the variation of film thickness with viscosity for a UWWC and UZZC. As expected, film thickness increases with increased viscosity for both cases. The predicted value of the film thickness matches well with the corresponding value computed from the Nusselt theory [34]. However, for ZZC, the film thickness also fluctuates. This may be due to the channel's zigzag shape. The film velocity accelerates at the tip and retards in the trough region to satisfy flow continuity. At higher viscosity, the range of fluctuation narrows. Note that the damping of turbulent eddies within the liquid film near both the liquid and solid surface is governed mainly by viscosity [35]. Subsequently, higher viscosity damps the oscillation.

Next, the film thickness is normalized by the viscous length scale as $\delta^* = \delta/(\nu_L^2/g)^{1/3}$. Figure 6(b) shows the variation of $\delta^*$ with Kapitza number for a UWWC. As expected, film thickness decreases with an increased Kapitza number. Note that a higher value of Kapitza number corresponds to a lower value of solvent viscosity. It also shows a scaling relation $\delta^* \sim Ka^{1/4}$. Earlier, Singh et al. [7] presented similar scaling for film falling over an inclined plate. Film thickness also has been reported to increase with solvent viscosity [7, 41], which is consistent with the Nusselt theory [34]. As mentioned earlier, film stability is governed by the competing effects between various forces acting on it. At high viscosity, enhanced viscous forces act against gravity and make the film more stable [16]. As a result, film thickness increases.

As already mentioned, flow simulations have been conducted for the oscillating inlet. Because of the oscillating inlet velocity, a wavy film appears. Wang et al. [27] observe that wavelength ($\lambda$) of the wavy film decreases with increased frequency ($f$) for OWWCs. The pronounced impact of the frequency is visible at $f = 60$ Hz. Accordingly, f = 60 Hz is chosen in the flow simulations for all viscosities. The wavelengths of the wavy film are computed and compared for both WWCs and ZZCs. The computed value of the wavelength also can be normalized as $\lambda^* = \lambda/(\nu_L^2/g)^{1/3}$. Figure 7 shows the variation $\lambda^*$ with Ka. As expected, the normalized wavelength increases with the increased value of Kapitza number for both cases. Note that the lower value of Kapitza number corresponds to higher viscosity. As noted, viscosity stabilizes the film and acts as damping factors for wave. Furthermore, the normalized wavelength $\lambda^*$ value is almost the same for both case, except slight discrepancy is observed at a higher value of Kapitza number.

### 4.2 Mass Transfer coefficient

In this section, we consider the mass transfer from gas to liquid phase due to physical absorption. The 2D simulation using the VOF method of co-current gas-liquid flow in the structure packed column is conducted to compute the interfacial mass transfer [24]. The mass transfer is not directly affected by the recirculation in the liquid side under corrugation cavities. Indeed, the chemical species only penetrate in a very thin concentration layer. Thus, the transfer process is controlled by the diffusion/advection at the film interface.

The methodology for computing the mass transfer coefficient is the same as in Wang et al. [27] and is only briefly presented here. The mass transfer coefficient ($k_L$) for physical transport via absorption is calculated using standard formula

$$k_L = j/\Delta p\,, \tag{35}$$

where $J$ is the mass transfer flux (mol/m$^2$·s) at the gas-liquid interface and $\Delta P$ is the driving force computed as:

$$\Delta P = \frac{P_{N_2O,in} - P_{N_2O,out}}{\ln\left(P_{N_2O,in}/P_{N_2O,out}\right)}. \tag{36}$$

Applying the ideal gas law, $\Delta P$ can be written in terms of the concentration of the $N_2O$ at the gas inlet and outlet:

$$\Delta P = \frac{\left(C_{N_2O,in} - C_{N_2O,out}\right)RT}{\ln\left(C_{N_2O,in}/C_{N_2O,out}\right)}, \tag{37}$$

where $R$ is the ideal gas constant (J/K·mol) and $T$ is the temperature in the unit of K. To compare the numerically computed value of the mass transfer coefficient with an analytically calculated one, the unit of the mass transfer is converted to *m/s*.

Figure 8 shows the variation of the mass transfer coefficients with viscosity for uniform inlet velocity (UWWC and UZZC) of $u_L = 0.17\ m/sec$. The UZZC shows higher mass transfer coefficient (especially at small viscosity) when compared to the WWC. This effect is due to the packing geometry and can be explained by the flow field in the packed column in Figure 9. Note that hydrodynamics plays a critical role in the mass transfer. In Figure 9, the recirculation zones are seen in the gas side for the UZZC. The recirculation zone enhances the convection. Previous studies by Haroun et al. [21] also uncovered a similar observation. The mass transfer is found to be higher for a triangular channel compared to a corresponding one for flat liquid film.

Figure 10 shows the variation of $k_L$ with solvent viscosity for both flat and wavy film flow. As expected, the mass transfer coefficient decreases with increased solvent viscosity in all four cases. Here, wavy film flow shows steep variation with viscosity. The difference between $k_L$ for wavy and smooth film is observed to be higher at lower viscosity. At higher viscosity, the difference becomes smaller. Increased value of viscosity damps the initial perturbation in the flow field. On the other hand, interfacial surface tension and gravitational forces dominate the viscous force at small viscosity. As a result, eddies at the interface occurs and promotes the mass transfer, in addition to higher interfacial area. In Figure 11(a), we compare the variation of the mass transfer coefficient with viscosity for expression (25) and numerically compute one at a fixed solvent and gas flow rate and wide range of viscosity ($\{\mu \sim 1 - 10\ cP$), where C= ??? Both plots match reasonably well, except the analytically derived expression shows slightly steep variation, i.e., higher exponent value (-0.36). In addition, the scaling for $k_L$ with solvent viscosity shows a scaling relation $k_L \sim {\nu_L}^{-0.34}$ for wavy film in the OWWC (see Figure 11(b)). This is nearly the same as the exponent value -0.36 in Eq. (25), as well as the experimental observation (-0.4 in Eq. (26)) by Song et al. [1, 12].

## 5. Conclusion

Post combustion carbon capture via solvent absorption in a packed column is a promising technology for mitigation of the greenhouse gas emissions. Absorption is carried out through countercurrent gas-liquid flow in a packed column. The mass transfer between the gas and liquid phases plays a critical role in determining the column height. Recent experiments have found the direct impact of viscosity on mass transfer is larger for a packed column ($k_L \sim \mu^{-0.40}$) compared to WWC ($k_L \sim \mu^{-0.17}$). In principle, this direct influence of viscosity on the mass transfer coefficient may be due to the liquid turbulence enhanced mass transfer in a packed column. In this context, detailed analytical studies and numerical simulations have been conducted to explain the underlying physics behind such discrepancy.

Wavy film flow is proposed as a major mechanism responsible for the discrepancy and expressions for eddy-enhanced diffusivity, and the mass transfer coefficient for wavy film falling over a flat plate is derived based on the fundamental hydrodynamic and mass transfer equations. It demonstrates a much higher dependence of mass transfer on viscosity and solvent flow rate in accordance to the experiment. In addition, to investigate the direct impact of viscosity on the mass transfer coefficient, multiphase flow simulation has been conducted for WWCs and ZZCs. ZZC was chosen for its 2D representation of a structured packed column. The predicted value of average film thickness ($\delta_{av}$) matches well with Nusselt theory. The value of $\delta_{av}$ increases with increased value of solvent viscosity. The wavelength of periodic wavy film decreases with increased viscosity. The WWC and ZZC show almost identical wavelength values, except at low viscosity where the WWC showed a slightly higher value. Notably, the mass transfer coefficient for a wavy film shows larger dependence on the viscosity compared to a flat uniform film, which is true for both WWC and ZZC. This dependence from simulation generally agrees with that ($k_L \sim \upsilon_L^{-0.35 \sim -0.4}$) found in both theoretical and experimental studies.

**Acknowledgments**

Pacific Northwest National Laboratory is operated by Battelle for the U.S. Department of Energy (DOE) under Contract No. DE-AC05-76RL01830. This work was funded by the DOE Office of Fossil Energy's Carbon Capture Simulation Initiative (CCSI) through the National Energy Technology Laboratory. Authors also acknowledge the valuable inputs and discussions of Prof. Gerry Rochelle (University of Texas at Austin).

## REFERENCE

1. Song, D., A.F. Seibert, and G.T. Rochelle, *Mass Transfer Parameters for Packings: Effect of Viscosity.* Industrial & Engineering Chemistry Research, 2018. **57**(2): p. 718-729.
2. Razi, N., O. Bolland, and H. Svendsen, *Review of design correlations for CO2 absorption into MEA using structured packings.* International Journal of Greenhouse Gas Control, 2012. **9**(0): p. 193-219.
3. Rochelle, G.T., *Amine Scrubbing for CO2 Capture.* Science, 2009. **325**(5948): p. 1652-1654.
4. Spiegel, L. and W. Meier, *Distillation Columns with Structured Packings in the Next Decade.* Chemical Engineering Research and Design, 2003. **81**(1): p. 39-47.
5. Mackowiak, J., *Fluid Dynamics of Packed Columns*. 2010, Berlin: Springer-Verlag.
6. Figueroa, J.D., et al., *Advances in $CO_2$ capture technology—The U.S. Department of Energy's Carbon Sequestration Program.* International Journal of Greenhouse Gas Control, 2008. **2**(1): p. 9-20.
7. Singh, R.K., J.E. Galvin, and X. Sun, *Three-dimensional simulation of rivulet and film flows over an inclined plate: Effects of solvent properties and contact angle.* Chemical Engineering Science, 2016. **142**: p. 244-257.
8. Whitman, W.G., *The two film theory of gas absorption.* International Journal of Heat and Mass Transfer, 1962. **5**(5): p. 429-433.
9. Higbie, R., *The rate of absorption of a pure gas into a still liquid during short periods of exposure.* Transactions of the American Institute of Chemical Engineers, 1935. **31**: p. 365-389.
10. Danckwerts, P.V., *Significance of Liquid-Film Coefficients in Gas Absorption.* Industrial & Engineering Chemistry, 1951. **43**(6): p. 1460-1467.
11. Biń, A.K., *Mass transfer into a turbulent liquid film.* International Journal of Heat and Mass Transfer, 1983. **26**(7): p. 981-991.
12. Song, D., A.F. Seibert, and G.T. Rochelle, *Effect of Liquid Viscosity on the Liquid Phase Mass Transfer Coefficient of Packing.* Energy Procedia, 2014. **63**: p. 1268-1286.
13. Tsai, R.E., et al., *Influence of viscosity and surface tension on the effective mass transfer area of structured packing.* Energy Procedia, 2009. **1**(1): p. 1197-1204.
14. Tsai, R.E., et al., *A dimensionless model for predicting the mass-transfer area of structured packing.* AIChE Journal, 2011. **57**(5): p. 1173-1184.
15. Wang, C., et al., *Packing Characterization for Post Combustion $CO_2$ Capture: Mass Transfer Model Development.* Energy Procedia, 2014. **63**: p. 1727-1744.
16. Ataki, A. and H.J. Bart, *Experimental and CFD simulation study for the wetting of a structured packing element with liquids.* Chemical Engineering & Technology, 2006. **29**(3): p. 336-347.
17. Fernandes, J., et al., *Application of CFD in the study of supercritical fluid extraction with structured packing: Wet pressure drop calculations.* The Journal of Supercritical Fluids, 2009. **50**(1): p. 61-68.
18. Hosseini, S.H., et al., *Computational fluid dynamics studies of dry and wet pressure drops in structured packings.* Journal of Industrial and Engineering Chemistry, 2012. **18**(4): p. 1465-1473.
19. Raynal, L., C. Boyer, and J.-P. Ballaguet, *Liquid Holdup and Pressure Drop Determination in Structured Packing with CFD Simulations.* The Canadian Journal of Chemical Engineering, 2004. **82**(5): p. 871-879.
20. Petre, C.F., et al., *Pressure drop through structured packings: Breakdown into the contributing mechanisms by CFD modeling.* Chemical Engineering Science, 2003. **58**(1): p. 163-177.
21. Haroun, Y., D. Legendre, and L. Raynal, *Direct numerical simulation of reactive absorption in gas–liquid flow on structured packing using interface capturing method.* Chemical Engineering Science, 2010. **65**(1): p. 351-356.

22. Haroun, Y., D. Legendre, and L. Raynal, *Volume of fluid method for interfacial reactive mass transfer: Application to stable liquid film.* Chemical Engineering Science, 2010. **65**(10): p. 2896-2909.
23. Marschall, H., et al., *Numerical simulation of species transfer across fluid interfaces in free-surface flows using OpenFOAM.* Chemical Engineering Science, 2012. **78**: p. 111-127.
24. Haroun, Y., L. Raynal, and D. Legendre, *Mass transfer and liquid hold-up determination in structured packing by CFD.* Chemical Engineering Science, 2012. **75**: p. 342-348.
25. Nieves-Remacha, M.J., L. Yang, and K.F. Jensen, *OpenFOAM Computational Fluid Dynamic Simulations of Two-Phase Flow and Mass Transfer in an Advanced-Flow Reactor.* Industrial & Engineering Chemistry Research, 2015. **54**(26): p. 6649-6659.
26. Wang, C., et al., *Beyond the standard two-film theory: Computational fluid dynamics simulations for carbon dioxide capture in a wetted wall column.* Chemical Engineering Science, 2018. **184**: p. 103-110.
27. Wang, C., et al., *Hierarchical calibration and validation for modeling bench-scale solvent-based carbon capture. Part 1: Non-reactive physical mass transfer across the wetted wall column.* Greenhouse Gases: Science and Technology, 2017. **7**(4): p. 706-720.
28. Hu, J., et al., *Numerical simulation of carbon dioxide ($CO_2$) absorption and interfacial mass transfer across vertically wavy falling film.* Chemical Engineering Science, 2014. **116**: p. 243-253.
29. Singh, R.K., et al., *Breakup of a liquid rivulet falling over an inclined plate: Identification of a critical Weber number.* Physics of Fluids, 2017. **29**(5): p. 052101.
30. Sebastia-Saez, D., et al., *3D modeling of hydrodynamics and physical mass transfer characteristics of liquid film flows in structured packing elements.* International Journal of Greenhouse Gas Control, 2013. **19**(0): p. 492-502.
31. Pigford, R.L., *Counter-Diffusion in a Wetted Wall Column*, in *Department of Chemistry and Chemical Engineering*. 1941, University of Illinois.
32. Pacheco, M.A., *Mass Transfer, Kinetics and Rate-Based Modeling of Reactive Absorption*, in *Chemical Engineering*. 1998, The University of Texas at Austin: Austin, TX.
33. Patnaik, V. and H. PerezBlanco, *Roll waves in falling films: An approximate treatment of the velocity field.* International Journal of Heat and Fluid Flow, 1996. **17**(1): p. 63-70.
34. Nusselt, W., *Die Oberflächenkondesation des Wasserdampfes.* Zeitschrift des Vereines Deutscher Ingenieure 1916. **60**(27): p. 541–546.
35. Yih, S.M., *Modeling heat and mass transfer in wavy and turbulent falling liquid films.* Wärme - und Stoffübertragung, 1987. **21**(6): p. 373-381.
36. Odier, P., et al., *Fluid mixing in stratified gravity currents: the Prandtl mixing length.* Physical Review Letters, 2009. **102**(13): p. 134504.
37. Shkadov, V.Y., *Wave flow regimes of a thin layer of viscous fluid subject to gravity.* Fluid Dynamics, 1967. **2**(1): p. 29-34.
38. Hirt, C.W. and B.D. Nichols, *Volume of fluid (VOF) method for the dynamics of free boundaries.* Journal of Computational Physics, 1981. **39**(1): p. 201-225.
39. CD-adapco *STAR-CCM+ 10.04 User Guide*. 2015.
40. Brackbill, J.U., D.B. Kothe, and C. Zemach, *A continuum method for modeling surface tension.* Journal of Computational Physics, 1992. **100**(2): p. 335-354.
41. Gu, F., et al., *CFD Simulation of Liquid Film Flow on Inclined Plates.* Chemical Engineering & Technology, 2004. **27**(10): p. 1099-1104.